**The use of kinematics to quantify gait attributes and predict gait scores in dairy cows**

Célia Julliot[1,4], Gabriel M. Dallago[2,4], Amir Nejati[1,4], Abdoulaye B. Diallo[3,4] and Elsa Vasseur [1,4]

[1]McGill University, Animal Science, Sainte-Anne-de-Bellevue, Québec, Canada,

[2] University of Manitoba, Animal Science, Winnipeg, Manitoba, Canada,

[3] Université du Québec à Montréal, Computer Science, Montréal, Québec, Canada,

[4] Research and Innovation Chair in Animal Welfare and Artificial Intelligence (WELL-E)

Comments: 27 pages, 3 figures, 5 tables

Category: Other Quantitative Biology (q-bio.OT)

Detecting walking pattern abnormalities in dairy cows early on holds the potential to reduce the occurrence of clinical lameness. This study aimed to predict gait scores in non-clinically lame dairy cows by using gait attributes based on kinematic data. Markers were placed on 20 anatomical landmarks on 12 dairy cows. The cows were walked multiple times through a corridor while recorded by six cameras, representing 69 passages. Specific gait attributes were computed from the 3D coordinates of the hoof markers. Gait was visually assessed using a 5-point numerical rating system (NRS). Due to the limited number of observations with NRS lower than 2 (n = 1) and higher than 3 (n = 6), the NRS labels were combined into three groups, representing NRS ≤ 2, NRS = 2.5, and NRS ≥ 3. The dataset was split into training and testing sets (70:30 ratio), stratified by the distribution of the NRS categories. Random forest (RF), gradient boosting machine (GBM), extreme gradient boosting machine (XGBM), and support vector machine (SVM) with a radial basis kernel models were trained using k-fold repeated cross-validation with hyperparameters defined using a Bayesian optimization. Accuracy, sensitivity, specificity, F1 score, and balanced accuracy were calculated to measure model performance. The GBM model performed best, achieving an overall accuracy and F1 score of 0.65 in the testing set. The findings of this study contribute to the development of an automated monitoring system for early identification of gait abnormalities, thereby enhancing the welfare and longevity of dairy cows.

1. Introduction

Lameness, characterized by altered gait resulting from pain or discomfort due to injury or disease of the hooves or legs [1], is a major health, welfare and economic concern in dairy farming. It is ranked as the third most common reason for involuntary culling [2] and the third most costly health issue in dairy cows, following mastitis and infertility [3]. Early detection of lameness is critical for effective management and improved animal welfare. However, producers often underestimate its prevalence, with diagnosis frequently occurring only in advanced stages when clinical signs become overt [4]. This delay in detection can exacerbate the condition, leading to prolonged pain for the animal and increased economic losses for the producer.

Early lameness detection is critical for effective management, necessitating the identification of subtle locomotion changes prior to the onset of clinical symptoms. Visual locomotion scoring systems assign scores based on observed gait characteristics [4, 5] and are commonly employed for lameness assessment. While these methods are relatively simple to implement, they are time-consuming and inherently subjective, requiring extensive training to ensure consistency and are often impractical for continuous monitoring of large herds [6]. Technological advancements offer the potential for more objective and sensitive lameness detection. Gait analysis technologies, including vision-based systems, wearable sensors, and pressure-sensitive walkways, enable continuous monitoring of individual gait parameters [7].

Kinematics, a subset of biomechanics, examines how a body moves through space and time without considering physical forces [8]. Applied to dairy cattle, kinematic analysis captures whole-body locomotion across multiple anatomical regions, offering a more detailed assessment of gait dynamics. These technologies offer detailed locomotion analysis, aiding in the detection of subtle gait abnormalities, enhancing animal welfare, and reducing economic costs [4]. However, early detection remains challenging across all detection modalities, with conceptual variations in its definition. Early lameness can be classified as detection occurring prior to observable clinical signs or before a producer recognizes the condition, including the automation of conventional gait assessment approaches [9]. Emerging technologies, such as computerized gait analysis via cameras, algorithms, sensors, and phone applications, offer promising avenues for strengthening on-farm lameness detection and management.

Machine learning algorithms are well adapted for processing high-dimensional kinematic data due to their ability to model complex interactions and non-linear relationships [10]. For instance, Karoui et al. [11] developed a binary classification model based on three-dimensional kinematic coordinates, achieving an accuracy of 90.8% and a precision of 92.2% in clinical lameness detection. However, further research is needed to refine these models beyond binary classifications, allowing for the identification of varying locomotor abilities and facilitating early intervention before lameness becomes clinically apparent. The integration of kinematics and machine learning, therefore, offers considerable potential to address the limitations of conventional gait assessments, providing a scalable and objective approach for early detection of lameness in dairy herds.

The objectives of this study were to 1) develop a methodology for obtaining gait attributes using kinematic raw data and assess their appropriateness, and 2) investigate links between these gait attributes and a 5-point numerical rating system (NRS) gait scoring to identify important early indicators of gait abnormality.

## 2. Material and methods

The use of animals in this project and all procedures were reviewed and approved by the Animal Care Committee of McGill University and affiliated hospitals and research institutes for the experimentations conducted at McGill University (protocol #2016–7794). All aspects of this study met the standards established by the Canadian Council on Animal Care to ensure the human and ethical use of animals in research.

## 2.1. Data collection

Gait kinematics data were collected on 12 lactating Holstein dairy cows (parity: 2.25 ± 1.09 and DIM: 143.33 ± 70.42) between January 18 to February 12, 2021. Reflective spherical markers (B & L Engineering, Santa Ana, Ca) were attached at 20 anatomical locations in the body of the cows. This was done by a trained person using a stencil (10cm x 10cm) to allow a consistent placement on different cows and days of data collection. Cows were led by a halter and walked in front of a recording system containing six high-performance cameras (Basler Ace, Ahrensburg, Germany). Three cameras positioned along each side of a 7-meter passageway corridor provided overlapping fields of view to ensure complete kinematic data acquisition. The recording system was calibrated once a day before video recording started. Each cow was walked and recorded multiple times to obtain as many passages as possible, totalling 69 passages, each one being three steps long. The videos of each passage were synchronized using the Vicon Motus capture system in the TEMPLO motion analysis software (CONTEMPLAS GmbH, Kempten, Germany) and later transferred to the Vicon Motus 3D motion analysis software (CONTEMPLAS GmbH, Kempten, Germany) to obtain the 3D coordinates and angles of each marker. The 3D coordinates were rotated around the YZ axis (transversal and vertical axis) to correct distortions at the beginning and at the end of each passage. Tracking graphs of time by space of the maker attached to the whither of the animals were used to estimate the rotation degree.

## 2.2. Gait attributes

A methodology was developed to measure gait attributes of distance (track-up and stride length), duration (stride time and stance time), velocity, and angle (joint flexion) based on 3D coordinates. Only the 3D coordinates of the markers attached to the coffin of each hoof and the angle of the markers attached to the hock joint of the rear legs were used to calculate gait attributes.

The track-up (cm) is the distance between consecutive hoofprints of the front and rear legs on the same side of the cow [12]. It was calculated using the differential distance in the longitudinal axis (disX; cm) and the combined longitudinal and transverse axes (disXY; cm) from the hoof markers' coordinates for each step on both sides of the cow (eq. 1 and eq. 2, respectively).

$$disX\ (cm) = XF_n - XR_n \tag{1}$$

$$disXY\ (cm) = \sqrt{((XF_n - XR_n)^2 + (YF_n - YR_n)^2)} \tag{2}$$

Where $XF_n$ corresponds to the 3D coordinates values when the front hoof hits the ground on the X axis on the nth step (n=1 to 3); $XR_n$ corresponds to the 3D coordinates values when the rear hoof hits the ground on the X axis on the nth step.

The stride length (cm) is the longitudinal displacement between two consecutive hoof strikes of the same hoof [13]. It was measured as the displacement of the hoof marker between consecutive strikes and calculated for each step (eq. 3).

$$SL\ (cm) = XON_n - XON(n+1) \tag{3}$$

Where $XON_n$ corresponds to the 3D coordinates when the hoof touches the ground for the nth step (n = 1 to 2) on the X axis.

The stride time (sec) is the interval between two consecutive strikes of the same hoof [13]. It was estimated as the time between consecutive steps using the hoof marker, calculated for each step individually (eq. 4).

$$ST = \frac{ON_n - ON(n+1)}{60} \tag{4}$$

Where $ON_n$ are the same as the stride length attribute, divided by 60 because the cameras have been calibrated at a frequency of 60 frames per second, so we can obtain a duration from the 3D coordinates.

The stance time (sec) is the interval a hoof remains in contact with the ground between strikes [12, 13]. It was estimated using the hoof marker for each step (eq.5).

$$Tm = \frac{ONn - OFFn}{60} \quad (5)$$

Where $ON_n$ corresponds to the 3D coordinates when the hoof touches the ground for the nth step (n = 1 to 3); $OFF_n$ corresponds to the 3D coordinates when the hoof does not touch the ground anymore for the nth step (n = 1 to 3).

The velocity (cm/sec) is the ratio of stride length to stride duration [13]. In this study, it was estimated as the ratio of the stride length distance (disXY, cm) to stride time (eq. 6 and eq. 7, respectively).

$$disXY = \sqrt{(XFn\ ON - XR(n+1)\ ON)^2 + (YFn\ ON - YF(n+1)ON)^2} \quad (6)$$

$$V\ (cm/sec) = \frac{disXY}{ST} \quad (7)$$

Where $XF_n$ ON corresponds to the 3D coordinate values of the front hoof in contact with the ground on the X axis at the nth step (n=1 to 2); $XR_n$ ON corresponds to the 3D coordinate values of the rear hoof in contact with the ground on the X axis on the nth step.

The joint flexion (degrees) is the extension and flexion degree of limb joints during walking [12]. In this study, it was measured as the angle of the hock joints using markers attached to the rear limbs (eq. 8).

$$JF = \frac{(Rn\ ON + Rn\ OFF + R(n+1)ON + R(n+1)OFF + R(n+2)ON + R(n+2)OFF}{6} \quad (8)$$

Where Rn ON corresponds to the 3D coordinates of the rear hoof when it contacts the ground for the nth step (n = 1 to 3); Rn OFF corresponds to the 3D coordinates of the rear hoof when it no longer contacts the ground for the nth step.

2.3. *Gait scoring*

Gait scoring was done for each video-recorded passage by a trained observer using the 5-point numerical rating system (NRS) proposed by Flower and Weary [14]. The NRS ranges from 1 to 5, in which 1 represents a healthy gait and 5 a severe lameness. If a cow exceeded the criteria of a given score but did not meet all the requirements of the next score, an interval of 0.5 was assigned. An observer was trained until obtaining intra-observer reliability of 90.9% based on the extended percentage agreement coefficient [15]. The same observer scored all video-recorded passages in a randomized order.

2.4. *Data analysis*

Statistical analysis was done using the R statistical software [16] and its specific packages.

2.4.1. Data Preparation

Based on the biological possibility and visual observation of our research team, gait attributes with biologically implausible values were flagged as outliers. Values higher than 33.97 cm and less than -33.92 cm for the track-up variable were considered outliers (n = 14). Similarly, stride length measurements above 186.87 cm (n = 5), velocity values over 195.36 cm per second (n = 4), and stance time values below 0.53 seconds (n = 4) were also considered non-plausible and treated as outliers. Flagged outliers were replaced with the passage average rather than being deleted to avoid loss of data and preserve overall data structure (i.e., three complete steps per passage).

The distribution of the NRS scores in our study is shown in Figure 1a. Due to the limited number of observations with NRS lower than 2 (n = 1) and higher than 3 (n = 6), the NRS labels were combined into three groups, representing NRS $\leq$ 2, NRS = 2.5, and NRS $\geq$ 3 (Figure 1b). This grouping approach was implemented as it aligns with our objective of early detection of gait abnormalities, which are represented by lower NRS values, allows enough observations in each of the levels to train predictive machine learning models.

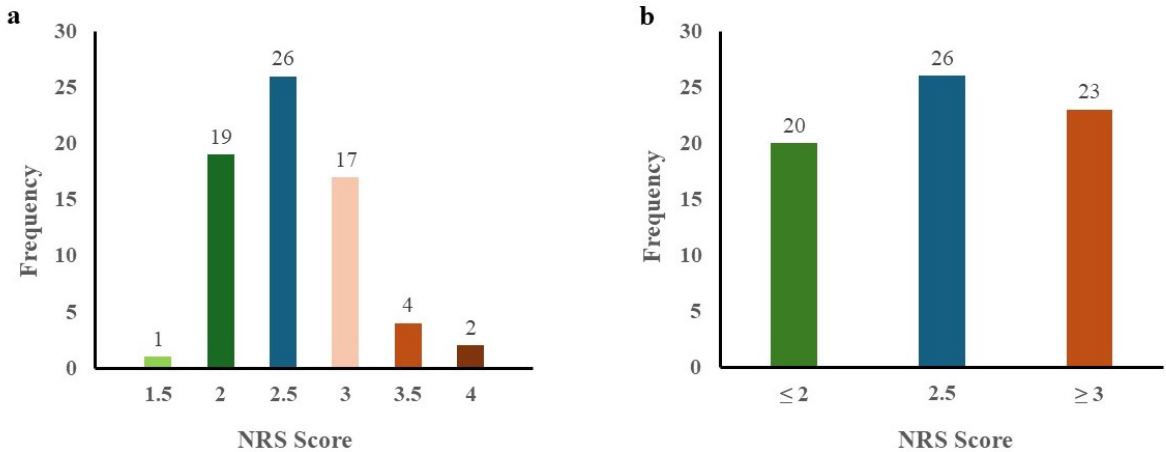

**Fig. 1.** Distribution of the numerical rating scale (NRS) values of dairy cows before (a) and after (b) grouping the scores into three categories (≤ 2, = 2.5, and ≥ 3).

The data was split into training (70%) and test (30%) sets based on the distribution of the combined NRS scores. The training set was used for model development and hyperparameter tuning, while the test set was held out for final predictive performance evaluation of the models. Because data were collected by passages, all observations belonging to a single passage were consistently assigned to either the training or test set to avoid information leakage.

2.4.2. Model training and evaluation

The algorithms random forest (RF), gradient boosting machine (GBM), extreme gradient boosting machine (XGBM), and support vector machine (SVM) with a radial basis kernel were evaluated to predict the combined NRS score based on the gait attributes. A repeated k-fold cross-validation was employed during training for hyperparameter optimization. To do that, the training set was further split into three folds (k=3) per repetition, with five total repetitions, ensuring that each fold preserved the distribution of the combined NRS scores. To avoid information leakage between folds, all observations belonging to a single passage were consistently assigned to the same fold. For the SVM model, the data were normalized to mean = 0 and standard deviation = 1 before model training. No normalization was done for the other models since tree-based models are not sensitive to differences in variable scales.

Bayesian optimization was used to identify the optimum hyperparameters for each of the models evaluated. The expected improvement acquisition function was used, with an epsilon set at 0.01 [17]. The number of iterations for the optimization process was set at 50. A minimum utility was set at 0.005, so the optimization process stopped in case the expected improvement was not greater than the minimum utility threshold. The following set of hyperparameters was evaluated:

- Random forest (*ranger* function from the *ranger* R package [18]): number of trees (50 to 200), number of variables to used to possibly split at in each node (1 to 12), minimal node size to split at (10 to 20), minimal terminal node size (10 to 20), and maximal tree depth (5 to 10).

- Gradient boosting machine (*gbm* function from the *gbm* R package [19]): number of trees (50 to 200), maximal depth of each tree (1 to 3), learning rate (0.001 to 0.1), and minimum number of observations in the terminal nodes (15 to 20).

- Extreme gradient boosting machine (*xgb.train* function from the *xgboost* R package [20]): maximum number of boosting iterations (50 to 300), maximum depth of a tree (3 to 6), learning rate (0.01, 0.1), subsample ratio of the training instance (0.6 to 0.8), subsample ratio of columns when constructing each tree (0.6 to 0.8), and minimum sum of instance needed in a split before considering further splits (5 to 15).

- Support vector machine with radial kernel filter (*svm* function from the *e1071* R package [21]): cost (0.0001 to 1) and gamma (0.001 to 1)

For each optimization iteration, a model with a set of hyperparameter combinations was trained using the 3-fold cross-validation repeated 5 times, each time evaluated based on a macro F1 score on the hold-out fold. An average F1 score per iteration was used to assess the predictive performance of the model. Once the optimal hyperparameters were identified, each algorithm was retrained on the full training set.

The models' predictive performance was evaluated based on overall accuracy and F1 score, as well as per-class sensitivity, specificity, balanced accuracy and F1 score. These metrics range from 0 to 1, and the closer the score is to 1, the better the model's predictive performance. The R code developed in this study is available on GitHub: https://github.com/CowLifeMcGill/Gait-prediction-Kinematic.git

2.4.3. Model interpretability

A two-step procedure was used to interpret the model with the best predictive performance using the DALEX R package [22]. First, global variable importance was evaluated through a permutation-based approach. A DALEX explainer was created using the *model_parts* function to quantify the dropout loss for each gait attribute, where a higher dropout loss indicated a greater importance of the attribute to the model's predictive performance. In the second step, accumulated local effect (ALE) plots were generated for the top-ranked important attributes using the *model_profile* function. ALE plots were employed to visualize the marginal effect of important attributes on the model's output relative to its average prediction.

## 3. Results

### 3.1. *Gait attributes*

Stride length and disXY track-up showed variations between the left and right legs of the cows (Table 1). On average, the right leg had a higher disXY distance (11.10 cm) than the left (10.09 cm), although both showed significant variability (SD ≈ 6 cm). The disX aspect of track-up revealed greater asymmetry (Table 1), with a positive mean value for the left leg (1.26 cm) and a slightly negative mean for the right leg (-0.09 cm). Stride length was generally greater on the right side for both front and rear legs, with the right rear leg showing the longest average stride (163.00 cm), closely followed by the right front leg (162.39 cm), the left rear leg (159.59 cm) and the left front leg (158.65 cm). Variability in stride length was similar between legs (SD ~10-12 cm), indicating consistent patterns within individuals but some inter-individual variation in locomotion.

Table 1. Descriptive statistics table for distance attributes for the final data to be analyzed. Attributes were measured on each side of the cow's body (right and left), on each leg (rear and front) or per passage.

| Attribute | Mean | SD | Min | Median | Max |
|---|---|---|---|---|---|
| Left leg disXY track-up (cm) | 10.09 | 6.02 | -15.11 | 10.13 | 33.97 |
| Right leg disXY track-up (cm) | 11.10 | 5.86 | 0.22 | 10.85 | 28.74 |
| Left leg disX track-up (cm) | 1.26 | 9.30 | -33.92 | 2.71 | 17.04 |
| Right leg disX track-up (cm) | -0.09 | 9.53 | -27.20 | 1.16 | 15.90 |
| Front left leg stride length (cm) | 158.65 | 11.84 | 124.31 | 159.09 | 185.06 |
| Rear left leg stride length (cm) | 159.59 | 10.96 | 127.31 | 160.58 | 180.33 |
| Front right leg stride length (cm) | 162.39 | 10.22 | 132.65 | 163.03 | 186.87 |
| Rear right leg stride length (cm) | 163.00 | 9.73 | 133.63 | 163.57 | 184.43 |

Stride and stance times were generally consistent for all four legs, with slight variations from one leg to the other (Table 2). The mean stride time ranged from 1.14 s (rear left leg) to 1.16 s (front right leg), with standard deviations of approximately 0.12–0.13 s, indicating relatively uniform timing patterns. Median stride times also varied minimally, suggesting symmetry in gait cycle durations across limbs. Stance times showed a similar consistency, with mean values ranging from 0.74 s (rear right leg) to 0.76 s (rear left leg). Although the overall variation was limited, rear legs tended to exhibit slightly longer stance times than front legs. The data also showed moderate individual variability (SD ~0.10–0.11 s), reflecting some natural differences in gait dynamics among cows.

Table 2. Descriptive statistics table for duration attributes for the final data to be analyzed. Attributes were measured on each side of the cow's body (right and left), on each leg (rear and front) or per passage.

| Attribute | Mean | SD | Min | Median | Max |
| --- | --- | --- | --- | --- | --- |
| Front left leg stride time (s) | 1.15 | 0.13 | 0.90 | 1.13 | 1.52 |
| Rear left leg stride time (s) | 1.14 | 0.12 | 0.90 | 1.12 | 1.48 |
| Front right leg stride time (s) | 1.16 | 0.13 | 0.93 | 1.15 | 1.48 |
| Rear right stride time (s) | 1.15 | 0.13 | 0.93 | 1.13 | 1.52 |
| Front left leg stance time (s) | 0.75 | 0.10 | 0.55 | 0.73 | 1.00 |
| Rear left leg stance time (s) | 0.76 | 0.11 | 0.53 | 0.75 | 1.12 |
| Front right leg stance time (s) | 0.75 | 0.10 | 0.55 | 0.73 | 1.03 |
| Rear right leg stance time (s) | 0.74 | 0.11 | 0.53 | 0.72 | 1.05 |

Leg velocity was relatively balanced across limbs, with slightly higher mean values on the right side (Table 3). The rear right leg exhibited the highest average velocity (144.62 cm/s), followed by the front right (142.94 cm/s), rear left (142.00 cm/s), and front left (140.70 cm/s) legs. Despite small inter-limb differences, standard deviations (~18–21 cm/s) indicated moderate variability across individuals. A minimum velocity of 0.00 cm/s was reported for the rear right leg, suggesting a potential measurement bias or moment of stance. Regarding joint angles (Table 3), the rear left leg had a slightly larger average angle (148.12°) compared to the rear right (145.40°), with somewhat more variability on the left (SD = 5.65° vs. 4.16°). These differences could reflect subtle asymmetries in leg kinematics during gait.

Table 3. Descriptive statistics table for speed and angle attributes for the final data to be analyzed. Attributes were measured on each side of the cow's body (right and left), on each leg (rear and front) or per passage.

| Attribute | Mean | SD | Min | Median | Max |
|---|---|---|---|---|---|
| Front left leg velocity (cm/s) | 140.70 | 19.38 | 89.21 | 136.41 | 190.06 |
| Rear left leg velocity (cm/s) | 142.00 | 18.15 | 96.09 | 140.65 | 195.35 |
| Front right leg velocity (cm/s) | 142.94 | 19.49 | 101.81 | 138.69 | 189.76 |
| Rear right leg velocity (cm/s) | 144.62 | 21.46 | 0.00 | 143.72 | 188.10 |
| Rear left leg average angle (°) | 148.12 | 5.65 | 141.20 | 146.49 | 175.45 |
| Rear right leg average angle (°) | 145.40 | 4.16 | 136.02 | 145.30 | 156.45 |

3.2. *Predictive model*

In the training dataset, the GBM model achieved the highest overall performance with an accuracy and F1-score of 0.93, outperforming the RF, XGBM, and SVM models (Table 4). Class-specific metrics for GBM were consistently high, with sensitivities ranging from 0.90 to 0.98, specificities from 0.92 to 1.00, and balanced accuracies from 0.94 to 0.95 across the three combined NRS categories. In contrast, the other models exhibited more variable performance; for instance, the RF model reached an overall accuracy of 0.73, while the XGBM and SVM models achieved overall accuracies of 0.80 and 0.70, respectively, with lower class-specific measures.

Table 4. Overall, and by NRS score categories, performance of random forest (RF), gradient boosting machine (GBM), extreme gradient boosting machine (XGBM), and support vector machine (SVM) models in predicting combined NRS scores based on gait attributes using the training data set. The model with the highest overall performance is shown in bold.

| Model | Overall | | Statistics by class | | | | |
|---|---|---|---|---|---|---|---|
| | Accuracy | F1-Score | NRS | Sensitivity | Specificity | Balanced Accuracy | F1-Score |
| RF | 0.73 | 0.72 | ≤ 2 | 0.52 | 0.91 | 0.72 | 0.60 |
| | | | 2.5 | 0.83 | 0.77 | 0.80 | 0.73 |
| | | | ≥ 3 | 0.81 | 0.92 | 0.87 | 0.83 |
| **GBM** | **0.93** | **0.93** | **≤ 2** | **0.90** | **0.97** | **0.94** | **0.92** |
| | | | **2.5** | **0.98** | **0.92** | **0.95** | **0.92** |
| | | | **≥ 3** | **0.90** | **1.00** | **0.95** | **0.95** |
| XGBM | 0.80 | 0.80 | ≤ 2 | 0.79 | 0.89 | 0.84 | 0.77 |
| | | | 2.5 | 0.83 | 0.88 | 0.86 | 0.81 |
| | | | ≥ 3 | 0.79 | 0.94 | 0.87 | 0.84 |
| SVM | 0.70 | 0.69 | ≤ 2 | 0.57 | 0.85 | 0.71 | 0.60 |
| | | | 2.5 | 0.81 | 0.79 | 0.80 | 0.74 |
| | | | ≥ 3 | 0.69 | 0.90 | 0.79 | 0.73 |

On the testing dataset (Table 5), overall F1 scores ranged from 0.58 to 0.65, with the GBM model showing the highest overall accuracy and F1 of 0.65. While the GBM model maintained high sensitivity for the ≤ 2 category (0.94), sensitivities for the 2.5 and ≥ 3 categories were lower across all models, suggesting challenges in discriminating these classes. Conversely, the XGBM model has the lowest overall accuracy (0.59) and F1 score (0.58).

Table 5. Overall, and by NRS score categories, performance of random forest (RF), gradient boosting machine (GBM), extreme gradient boosting machine (XGBM), and support vector machine (SVM) models in predicting combined NRS scores based on gait attributes using the testing data set. The model with the highest overall performance is shown in bold.

| Model | Overall | | Statistics by class | | | | |
|---|---|---|---|---|---|---|---|
| | Accuracy | F1-Score | NRS | Sensitivity | Specificity | Balanced Accuracy | F1-Score |
| RF | 0.60 | 0.60 | ≤ 2 | 0.78 | 0.80 | 0.79 | 0.68 |
| | | | 2.5 | 0.46 | 0.85 | 0.65 | 0.54 |
| | | | ≥ 3 | 0.62 | 0.76 | 0.69 | 0.59 |
| **GBM** | **0.65** | **0.65** | **≤ 2** | **0.94** | **0.87** | **0.91** | **0.83** |
| | | | **2.5** | **0.46** | **0.87** | **0.67** | **0.55** |
| | | | **≥ 3** | **0.62** | **0.74** | **0.68** | **0.58** |
| XGBM | 0.59 | 0.58 | ≤ 2 | 0.89 | 0.76 | 0.82 | 0.71 |
| | | | 2.5 | 0.46 | 0.87 | 0.67 | 0.55 |
| | | | ≥ 3 | 0.48 | 0.76 | 0.62 | 0.49 |
| SVM | 0.62 | 0.61 | ≤ 2 | 1.00 | 0.78 | 0.89 | 0.78 |
| | | | 2.5 | 0.38 | 0.87 | 0.62 | 0.47 |
| | | | ≥ 3 | 0.57 | 0.79 | 0.68 | 0.57 |

3.3. *Model interpretability*

The variables that had the greatest influence on the GBM model's predictions are ranked from most influential to least influential in Figure 2. Angle and track-up attributes were the most influential variables when predicting cow locomotion score. Conversely, velocity and stance time were less influential.

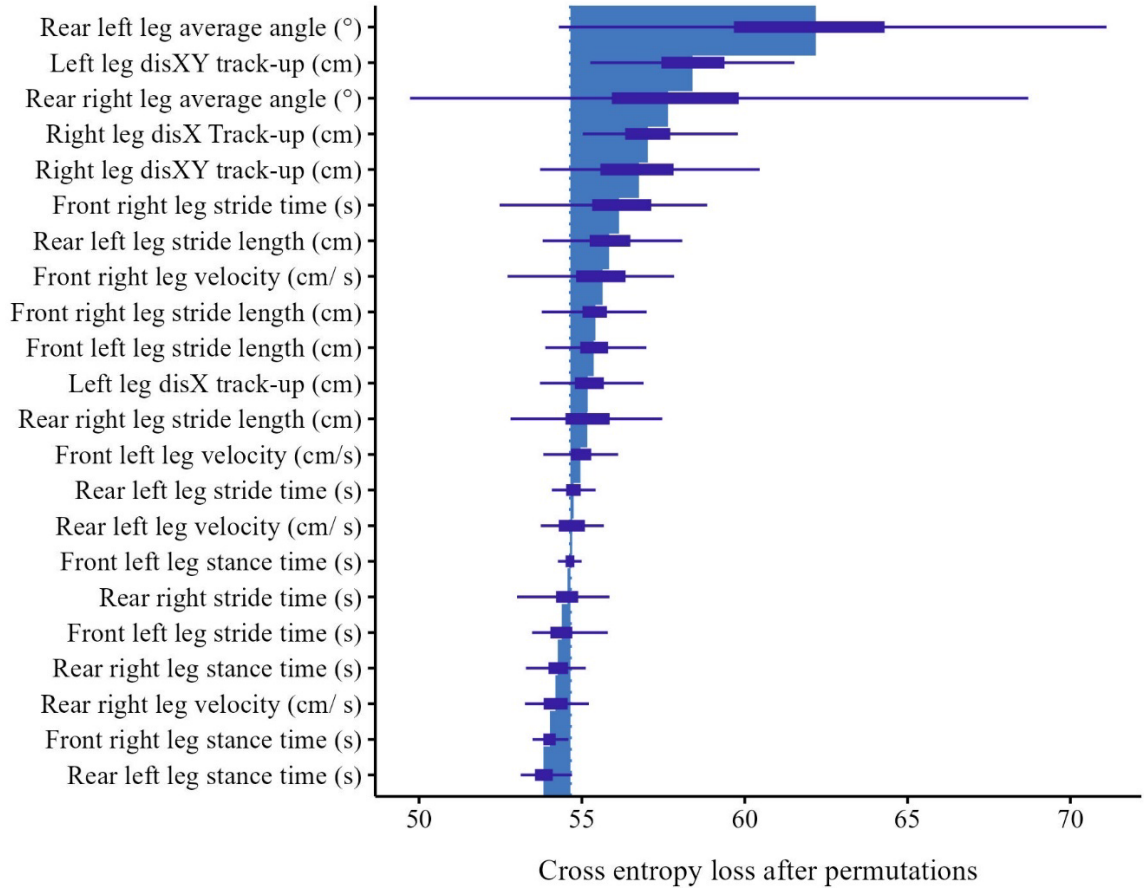

**Fig. 2.** Permutation-based feature importance (cross-entropy loss) for the GBM model, which had the best predictive performance. Each bar shows the average cross-entropy loss after randomly permuting the values of a given predictor 100 times. Larger increases in cross-entropy loss indicate a greater contribution of that predictor to the model's predictive accuracy. Box plots represent the distribution across permutation iterations.

Figure 3 shows the shape of the relationship between the top six most important gait attributes on model predictions. It can be seen that a high value on the left leg angle attribute was associated with an NRS score ≤ 2. Conversely, a high value on the stride time attribute is associated with an NRS score ≥ 3. In another sense, high track-up values are associated with NRS = 2.5 and NRS ≥ 3.

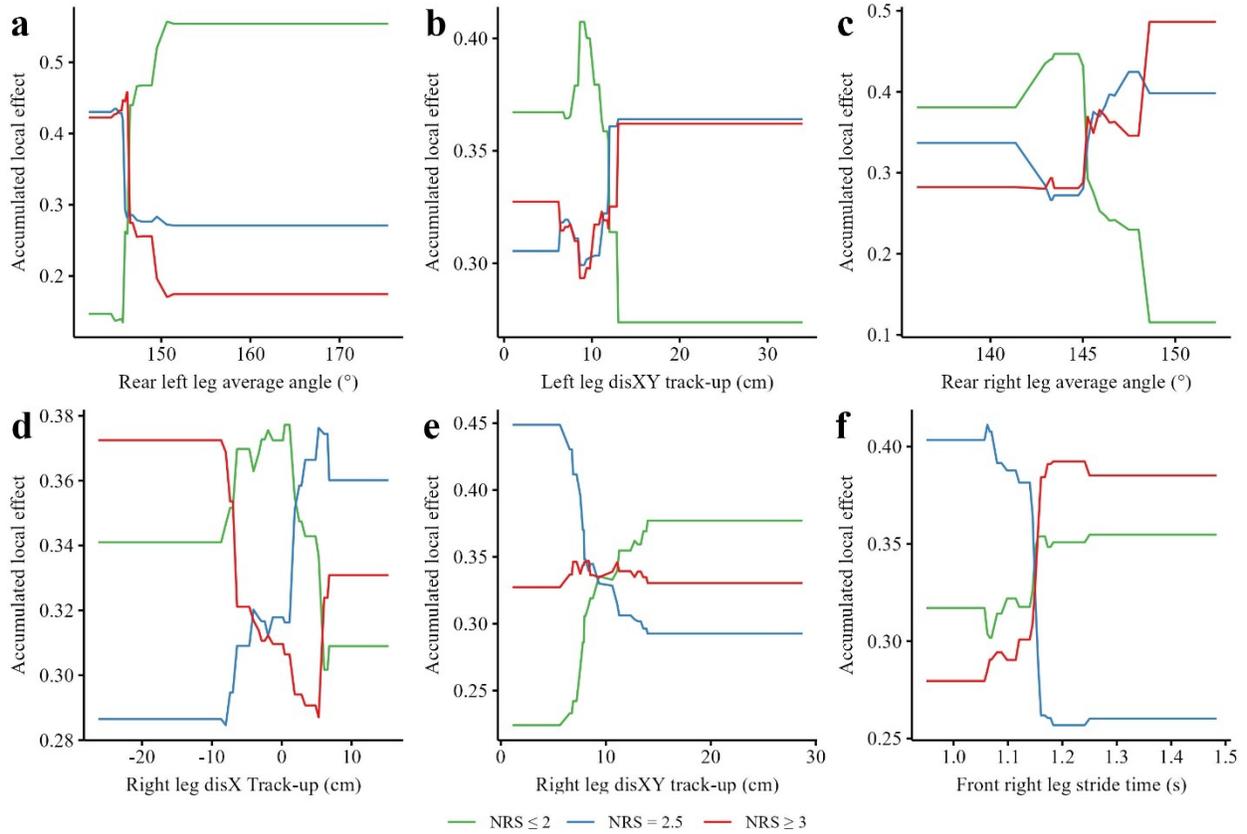

**Fig. 3.** Accumulated local effect (Y axis) plots for the six most important predictors (X axis). Each line represents how the predicted probability of each combined NRS category (NRS ≤ 2, NRS = 2.5, and NRS ≥ 3) changes relative to the average prediction as the predictor variable increases. Positive deviations suggest an increase in predicted probability, whereas negative deviations indicate a decrease.

## 4. Discussion

Lameness is a critical welfare and productivity concern in dairy production systems, justifying the need for objective and early detection methods. Quantifying gait attributes from 3D kinematic coordinates reduces the subjectivity associated with visual assessments and allows detailed temporal and spatial analyses of locomotion [7]. This study aimed to develop a methodology for extracting gait-specific attributes and evaluated their relevance for identifying early gait abnormalities in dairy cows. The gait attributes included in this study were selected based on previous studies identifying key indicators for characterizing gait in dairy cattle [13, 14, 23]. As gait expression varies across cows, whether classified as normal or abnormal, integrating multiple attributes such as joint angle, track-up, gait symmetry and stride time allows prediction models to better capture individual locomotor variability [5]. This integration is particularly important because cows exhibit inherent variability in movement across time and space. Predictive models built on multiple quantitative variables may therefore be more informative in the gait expression diversity than models relying directly on raw 3D coordinate data for direct prediction of NRS [24].

Previous automated systems for lameness detection, using deep learning, have primarily relied on binary classification models to detect a clinical case of lameness versus non-lame. Wu et al. [23] developed a binary classification model for lame versus non-lame cows using stride length as the sole input variable, achieving 98.57% accuracy. Karoui et al. [11] utilized 3D kinematic data from 16 body markers to develop a deep-learning model based on leg joint movement curves, attaining a classification accuracy of 90.76%. Building upon these studies, Bradtmueller [24] advanced the Karoui et al. [11] model by employing a neural network to classify cows on a 5-point NRS scale [14]. This study represented the first attempt to integrate full-body 3D motion tracking with deep learning for lameness assessment. Building on the results proposed by Bradtmueller [24], our study developed a model aimed at improving the identification of cow-specific attributes that can serve as reliable indicators of gait abnormalities. The resulting model highlighted track up, joint flexion and stride time as potentially informative parameters for detecting gait abnormalities.

The performance of the GBM model provides important insights for early detection. Although the overall F1-score was moderate (0.65), the model performed very strongly for cows with low NRS scores ($\leq 2$), achieving a sensitivity of 0.94 and balanced accuracy of 0.91. This result aligns closely with our primary objective, which focused on detecting subtle gait deviations in cows that are not clinically lame. In contrast, performance was lower for intermediate and high scores (NRS 2.5 and ≥ 3), which is expected given the limited number of cows in these categories and the narrow distribution of scores in our dataset (N = 12). The model's strong predictive ability for low scores suggests that quantitative gait attributes can detect early deviations before they are fully captured by subjective scoring systems.

In our GBM model, joint angle and track-up were the most influential gait attributes for lameness classification, whereas velocity and stance time attributes had comparatively lower importance. The detection of abnormal gait relies on a comprehensive assessment of locomotor characteristics [4]. Although the relationship between specific gait attributes and lameness severity is still not fully understood, several studies provide evidence of relevant patterns. For instance, Pluk et al. [25] demonstrated that lameness in dairy cows is associated with measurable alterations in fetlock joint kinematics. Using an automated vision-based system, they quantified both initial ground contact and lift-off angles, as well as the range of motion of the fore- and hindlimb fetlock joints. Their results indicated that, compared with non-lame cows, lame cows exhibited significantly increased release angles and reduced range of motion in the forelimbs, as well as altered touch angles in the hindlimbs. This suggests that the presence of modifies the movement patterns of the limbs, highlighting the potential of joint-angle measurements as objective indicators for early detection of locomotor disorders in dairy cattle. Similarly, Telezhenko et al. [26] found that lame cows walked more slowly and with shorter stride lengths than non-lame cows. On hard concrete floors, lame cows adopted a wider step angle, likely as compensation for discomfort; this difference disappeared when cows walked on more compliant floors (rubber mats or sand). Maertens et al. [27] noted that lame cows tended to have shorter strides length compared with non-lame cows, reflecting compensatory adjustments to reduce discomfort or pain in the limbs. The highlighted that stride length, in combination with other gait variables, could effectively distinguish lame from non-lame cows under experimental conditions. While studies on stiff joint movement are limited, prior research has suggested that joint flexion angle may serve as a relevant lameness indicator [25]. Track-up showed greater variability compared to other variables (from -15 to 34 cm on the left and 0.22 – 29 on the right, Fig 2 and Fig 3), which could indicate greater capacity to distinguish the NRS scores explaining why this variable was considered important for predictions. A notable pattern was that left-side track-up and angle attributes were more influential than right-side values. However, we did not record leg-specific clinical injuries.

A proposed explanation concerns anatomical asymmetry, although direct causal proof is lacking. Left-sided abomasal/visceral disorders are common in dairy cows, which demonstrates that substantial asymmetric abdominal contents and associated postural changes occur in production animals [28]. Retrospective clinical analyses also show that lameness and claw lesions are frequently diagnosed in cows presenting with left-displaced abomasum, although the available evidence does not indicate a consistent left–right predominance of claw lesions in these cases [29]. Independent experimental work demonstrates that cows redistribute their body weight and reduce loading on a painful limb when experiencing hoof discomfort [30], so any chronic asymmetric abdominal mass or altered posture that subtly shifts limb loading could plausibly increase left-side limb stress over time. Furthermore, dairy cows exhibit various forms of behavioral lateralization, including biases in visual inspection and posture, but these patterns are variable and not systematically associated with limb health [31, 32]. Collectively, these findings confirm that cattle display both anatomical and behavioral asymmetries, while indicating that current evidence does not establish a directional asymmetry in limb disorders. The hypothesis is biomechanically plausible and warrants investigation.

Quantifying gait attributes using 3D kinematic data offers significant advantages over subjective locomotion scoring. Conventional NRS classifications depend on observer interpretation, which can lead to inconsistencies, particularly when cows with different gait abnormalities score the same NRS. In contrast, digital quantification of gait parameters provides a more accurate temporal and spatial assessment of locomotor patterns, reducing observer bias [5, 7]. Quantified data provides a dynamic, longitudinal view of each animal's locomotor status, rather than static snapshots offered by NRS scoring, and allows models to learn individual baseline patterns for personalized risk assessments based on deviations from a cow's own norm. Early-stage lameness may not cause an observable change in categorical scoring systems (such as moving from an NRS of 1 to 2). However, quantified gait metrics can capture early dysfunctions (for example, a 5% increase in stride time variability). Quantified gait attributes are more universal than subjective scores, which often vary between observers or farms. For example, a stride length of 1.10 meters is objectively shorter than 1.30 meters, independent of interpretation. From a predictive model perspective, quantified gait attributes can improve the informativeness and utility of training data. Continuous data carries more detailed information than categorical classes, allowing models to detect subtle gait changes that preceded over lameness. They allow models to explore complex relationships between variables (e.g., a combination of shorter stride time and higher track-up predicting increased risk) instead of being restricted by fixed scoring categories like NRS and reducing labelling noise, leading to more robust and reliable models.

Early detection of lameness and subclinical leg pathologies is of growing interest in both biological and computational research. A model capable of predicting subtle gait changes can significantly improve economic benefits by improving pathology management and enhancing treatment efficacy [6]. As herd sizes increase, the implementation of automated lameness detection systems is becoming a necessity to support management decisions, rather than a research question [27]. While our models are not yet ready for farm implementation, they contribute to a detailed understanding of gait attributes, enhancing our understanding of cow gait abnormalities and the causes of leg issues. One of the main limitations of this study is the narrow range of NRS scores due to the small sample size of dairy cows (N=12) and the uneven distribution of scores. To improve predictive accuracy and robustness, future models will need to incorporate a broader and more diversified range of locomotion scores by including more cows with varied walking patterns. This will enable the development of a more detailed model that captures greater inter- and intra-individual differences in gait and potentially improves the ability to predict early lameness.

## 5. Conclusions

This study developed a methodology, based on machine learning model, for obtaining gait-specific attributes from 3D kinematic coordinates and evaluated their relevance for early detection of gait abnormalities in dairy cows. The best predictive model achieved an overall accuracy of 65%, with particularly strong performance in detecting cows with low NRS scores (NRS $\leq 2$), the stage most relevant for early intervention. Joint angle and track-up emerged as the most influential gait attributes, while velocity and stance-time attributes contribute the least. These findings demonstrate the value of gait attributes for identifying early locomotor changes and highlight the potential of quantitative kinematic analysis to complement traditional locomotion scoring. Future studies should incorporate larger datasets with a broader distribution of lameness severity to refine model performance and improve the capacity for precision detection of emerging locomotor problems.

**Funding**

Funding for conducting research and student stipend support for C. J. was provided by Vasseur's Established Researcher NSERC Discovery Grant (ID# RGPIN-2019-04728). Equipment for the kinematic platform used in the study was funded through Vasseur's Canada Foundation for Innovation's John R. Evans Leaders Fund (ID# JELF-20258). Additional stipend and salary funding to G. M. D and A. N. was provided through Vasseur and Diallo's Research and Innovation Chair in Animal Welfare and Artificial Intelligence (WELL-E) established with funding from Natural Sciences and Engineering Research Council of Canada (NSERC Alliance grant; ID# ALLRP 570894-2021), PROMPT and industrial partners Novalait, Dairy Farmers of Canada, Dairy Farmer of Ontario, Les Producteurs de lait du Québec, Lactanet.

**Declaration of Competing Interest**

The authors declare that they have no known competing financial interests or personal relationships that could have appeared to influence the work reported in this paper.

**CRediT Authorship Contribution statement**

**Célia Julliot**: Conceptualization, Formal Analysis, Investigation, Methodology, Software, Visualization, Writing – Original Draft, Writing – Review & Editing. **Gabriel M. Dallago**: Conceptualization, Data Curation, Formal Analysis, Investigation, Methodology, Software, Visualization, Writing – Original Draft, Writing – Review & Editing. **Amir Nejati**: Conceptualization, Data Curation, Formal Analysis, Investigation, Methodology, Software, Writing – Original Draft, Writing – Review & Editing. **Abdoulaye B. Diallo**: Conceptualization, Funding Acquisition, Methodology, Project Administration, supervision, Writing – Review &

Editing. **Elsa Vasseur**: Conceptualization, Funding Acquisition, Methodology, Project Administration, Supervision, Writing – Original Draft, Writing – Review & Editing.

## Data availability statement

The data that has been used is confidential. But the R code used to develop the prediction models are available on GitHub: https://github.com/CowLifeMcGill/Gait-prediction-Kinematic.git